\documentclass[twocolumn, aps, superscriptaddress,pdflatex]{revtex4-2}
\usepackage[english]{babel}
\usepackage[utf8]{inputenc}
\usepackage{lmodern}
\usepackage{dcolumn}
\usepackage{amssymb, amsmath}
\usepackage{graphicx}
\graphicspath{{figures/}} 
\usepackage{bm}
\usepackage{xcolor}
\usepackage{nicefrac}
\usepackage{hyperref}
\usepackage{booktabs}
\usepackage{multirow}
\usepackage{microtype}
\usepackage{braket}
\usepackage{physics}
\usepackage{changes}
\usepackage{lineno}

\begin{document}

\title{Ab-initio study of the beta Fe2O3 phase}

\author{Priyanka Mishra, Carmine Autieri}

\affiliation{\quad International Research Centre Magtop, Institute of Physics, Polish Academy of Sciences,
Aleja Lotnik\'ow 32/46, PL-02668 Warsaw, Poland; autieri@magtop.ifpan.edu.pl}

\begin{abstract}
We present first-principles results on the electronic and magnetic properties of the cubic bulk $\beta$-phase of iron(III) oxide (Fe$_2$O$_3$). Given that all Fe-Fe magnetic couplings are expected to be antiferromagnetic within this high-symmetry crystal structure, the system may exhibit some signature of magnetic frustration, making it challenging to identify its magnetic ground state.
We have analyzed the possible magnetic phases of the $\beta$-phase among which there are ferrimagnets, altermagnets and Kramers antiferromagnets. While the $\alpha$-phase is an altermagnet and the $\gamma$-phase is a ferrimagnet, we conclude that the magnetic ground state for the bulk $\beta$-phase of Fe$_2$O$_3$ is a Kramers antiferromagnet, moreover, we find that close in energy there is a bulk d-wave altermagnetic phase. We report the density of states and the evolution band gap as a function of the electronic correlations, for suitable values of the Coulomb repulsion the system is a charge-transfer insulator with an indirect band gap of 1.5 eV. More in detail, the unit cell of the $\beta$-phase is composed of 8 Fe$_a$ atoms and 24 Fe$_b$ atoms. The 8 Fe$_a$ atoms lie on the corner of a cube and their magnetic ground state is an G-type. This structural phase is composed of zig-zag chains Fe$_a$-Fe$_b$-Fe$_a$-Fe$_b$ with spin configuration $\uparrow$-$\uparrow$-$\downarrow$-$\downarrow$ along the 3 directions such that for every Fe$_a$ atoms there are 3 Fe$_b$ atoms. As the opposite to the $\gamma$-phase, the magnetic configuration between first-neighbor of the same kind is always antiferromagnetic while the magnetic configuration between Fe$_a$ and Fe$_b$ is ferro or antiferro. In this magnetic arrangement, first-neighbor interactions cancel out in the mean-field estimation of the N\'eel temperature, leaving second-neighbor magnetic exchanges as the primary contributors, resulting in a N\'eel temperature lower than that of other phases. Our work paves the way toward the ab initio study of nanoparticles and alloys for the $\beta$-phase of Fe$_2$O$_3$.
\end{abstract}

\maketitle

\section{Introduction}

The $\alpha$-phase of iron(III) oxide (Fe$_2$O$_3$) is commonly known as hematite, a naturally occurring mineral. It is the most stable polymorph of Fe$_2$O$_3$ under standard conditions. Hematite has a rhombohedral crystal structure, belonging to the hexagonal crystal system, and is typically found in nature as reddish-brown crystals or powder.
This phase has high chemical stability and hardness, making it useful in pigments, coatings, and as a raw material in iron production. It is also studied for potential applications in high-performance photocatalysis\cite{molecules28186671}, electrocatalysts for an enhanced hydrogen evolution reaction\cite{molecules29133082} and spintronics due to its semiconductor-like properties. The $\alpha$-phase (as bulk, nanoparticle and nanotube structures) is widely researched in material science due to its abundance and functional properties in nanotechnology.\cite{D0NR02705G,PhysRevB.109.224408,PhysRevB.109.195203}
Fe$_2$O$_3$ nanoparticles can also act as an efficient and recyclable heterogeneous catalyst for heavy metal adsorption\cite{molecules29194583}.

In terms of magnetic properties, $\alpha$-Fe$_2$O$_3$ is a centrosymmetric altermagnet below its N\'eel temperature (around 956 K) and exhibits weak ferromagnetism due to a canting of the spins caused by staggered 
Dzyaloshinskii–Moriya interaction
\cite{DZYALOSHINSKY1958241,galindezruales2024altermagnetismhoppingregime,autieri2023dzyaloshinskii,PhysRevB.106.L220404}. 
The altermagnetic order describes an antiferromagnetic system where the zero net magnetization in the non-relativistic spin limit is guaranteed by the spin-up and spin-down sublattices connected by rotation (proper or improper and symmorphic or nonsymmorphic), but the system also hosts the breaking of time-reversal symmetry with a non-relativistic spin-splitting in the band structure.\cite{Smejkal22,D3NR03681B,Cuono23orbital,grzybowski2024wurtzite,Fakhredine23} 
The $\alpha$-Fe$_2$O$_3$ magnetic configuration ($\uparrow\downarrow\uparrow\downarrow$) is the only magnetic configuration of the space group R$\overline{3}$C (no. 167) which produce altermagnetism.\cite{galindezruales2024altermagnetismhoppingregime,verbeek2024nonrelativisticferromagnetotriakontadipolarorderspin}
However, other magnetic configurations of the space group R$\overline{3}$C with stoichiometry A$_2$O$_3$ can produce altermagnetism and weak ferromagnetism if we add doping and/or external electric field to lower the symmetry.\cite{Bousquet_2024}

The $\gamma$-phase of iron(III) oxide ($\gamma$-Fe$_2$O$_3$), also known as maghemite, is a metastable polymorph of Fe$_2$O$_3$. It has a cubic spinel structure and exhibits ferrimagnetic properties, making it useful in magnetic recording and data storage. Maghemite can be synthesized by oxidizing magnetite (Fe$_3$O$_4$) and is often used in nanoparticle form for biomedical applications, such as drug delivery and magnetic resonance imaging (MRI).\cite{nano12213873} Though metastable, $\gamma$-Fe$_2$O$_3$ can transform into the more stable $\alpha$-phase at high temperatures.\cite{Danno2013}
The $\beta$-phase of iron(III) oxide ($\beta$-Fe$_2$O$_3$) is a rare and metastable polymorph and typically forms under specific high-temperature and pressure conditions. The $\beta$-phase can coexist with the $\alpha$-phase\cite{molecules28155722}, therefore, studies on the $\beta$-phase are essential to differentiate the origin of the chemical and physical effects arising from both phases. 
Structurally, the $\beta$-Fe$_2$O$_3$ phase is cubic and adopts a body-centered crystal structure of bixbyite type with Ia$3^-$ space group and a lattice parameter \textit{a} = 9.56 {\AA}. The structure features two kinds of inequivalent FeO$_6$ octahedra each characterized by specific patterns. These inequivalent FeO$_6$ octahedra are interconnected through a combination of edge-sharing and corner-sharing. \cite{Danno2013}
Experimentally, it exhibits antiferromagnetism with a N\'eel temperature of 119 K\cite{Tuek2015} which is much lower with respect to the $\alpha$-phase. 
Unlike the more common $\alpha$ and $\gamma$-phases, $\beta$-Fe$_2$O$_3$ is less studied due to its instability and difficulty in synthesis. Its unique structure gives it distinct magnetic and electronic properties, which are of interest in fundamental research. However, $\beta$-Fe$_2$O$_3$ is not commonly found in nature due to its structural instability that affects its chance of formation.
Both $\gamma$-Fe$_2$O$_3$ and $\beta$-Fe$_2$O$_3$ are metastable phases of iron oxide. At the nanoscale, they both undergo phase transformations as the particle size decreases.\cite{Sakurai2009} Since the $\gamma$ and $\beta$-phases have both cubic phases, the same stoichiometry and the same electronic configuration, we expect that the magnetic properties would be similar, however, one case is ferrimagnetic and the other has zero net magnetization. In this paper, we will highlight the difference between these two phases.
To our knowledge, no density functional theory calculations have been provided for the $\beta$-phase of Fe$_2$O$_3$. In this paper, we fill this research gap in the literature.\\

The paper is organized as follows: in the next Section, we
describe the computational details of the ab initio calculations.
In Sec. 3 we search for the magnetic ground state. In Sec 4, we present our results for the magnetic properties, electronic band structure and related density of states (DOS) for the magnetic ground state.
Finally, the last Section is devoted to our final remarks and conclusions.\\

\begin{figure}
    \includegraphics[width=\linewidth]{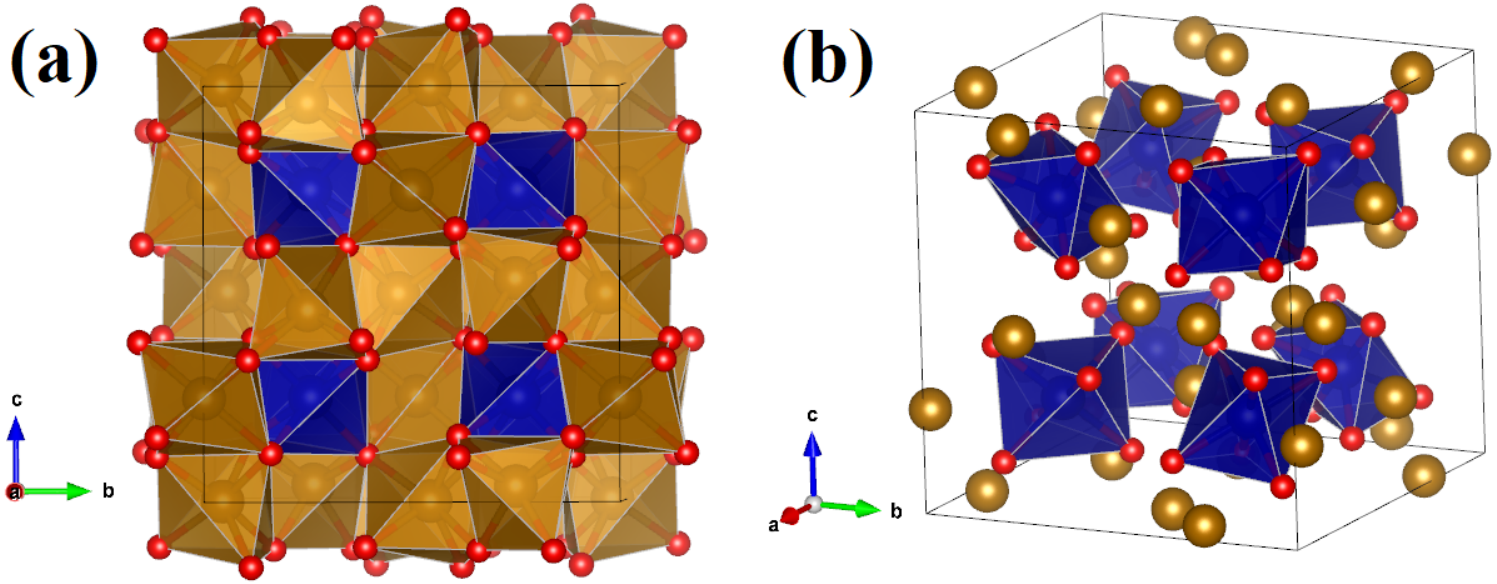}
    \caption{Crystal structure of the $\beta$-Fe$_2$O$_3$ phase. The Fe$_a$, Fe$_b$ and oxygen atoms are represented by blue, brown and red balls, respectively. (a) The system presents two kinds of FeO$_6$ octahedra which are Fe$_a$O$_6$ and Fe$_b$O$_6$ represented in blue and brown, respectively. (b) There are 8 Fe$_a$O$_6$ octahedra which are centered at the high-symmetry positions (0.50$\pm$0.25a,0.50$\pm$0.25a,0.50$\pm$0.25a) where a is the lattice constant.}
    \label{crystal}
\end{figure}

\section{Theoretical methods}

The first-principles calculations of the atomic and electronic structures are performed based on density functional theory (DFT)\cite{PhysRev.136.B864} as implemented in the Vienna \emph{ab initio} simulation package (VASP) based on the plane-wave basis set\cite{Kresse93,Kresse96b,Kresse99}.
The pseudopotentials are described using the Projector Augmented Wave (PAW) method, and the exchange-correlation functional is treated within the generalized gradient approximation (GGA) developed by Perdew-Burke-Ernzerhof (PBE). \cite{Perdew96}
We have used a PAW pseudopotential with 8 valence electrons for the Fe (4s$^1$3d$^7$) and 6 valence electrons for the O (2s$^2$ 2p$^4$). 
The cutoff energy of 530 eV was applied for the plane-wave expansion. The criterion to stop the self-consistent calculation was $10^{-5}$ eV/atom. The crystal properties were extracted from the Materials project website\cite{Jain2013} using the data of the mp-565814 optimized structure. The irreducible Brillouin zone was sampled using an 8$\times$ 8 $\times$ 8  k-point mesh centered in $\Gamma$ for the bulk $\beta$-Fe$_2$O$_3$. We have carried out the convergence of energy cutoff and checked the Brillouin zone meshing. 	
The Liechtenstein approach\cite{PhysRevB.52.R5467} was employed to evaluate the strong correlation effect on the 3$d$ orbital, with the correlation strength represented by the effective Hubbard $U$ and Hund coupling J$_H$ on the $3d$ Fe orbitals. We scan the value of U from 0 to 6 eV keeping J$_H$=0.15U which is a typical value for transition metals. In the literature, the Coulomb repulsion that was more used for the different polymorphs of the Fe$_2$O$_3$ is U=4 eV\cite{PhysRevB.69.165107,NAVEAS2023106033}.
For the Figures containing the crystal structure, we have used the visualization software VESTA\cite{Momma:db5098}.

\begin{table*}
    
    \begin{tabular}{|c|c|c|c|c|} \hline 
         Magnetic configuration & U=0   &  U=2   & U=4 & U=6\\ \hline 
            E$_{FM}$   &  8.63650 & 6.46846 &  4.49517 & 3.20209 \\  \hline 
          E$_{FiM}$   &  1.80028 & 1.31146 &  0.94402 & 0.67864 \\  \hline 
 F-type +(FM,FM,FM) &  0.09256 & 0.14929 &  0.14239 & 0.12073 \\  \hline 
 G-type +(FM,FM,FM)    &  0 & 0 &  0 & 0 \\  \hline 
    \end{tabular}
    \caption{Total energy values in eV for the bulk $\beta$-Fe$_2$O$_3$ for 4 different magnetic configurations as a function of the U values. The first and second rows report the results for the ferromagnetic and ferrimagnetic configurations, respectively. The third row reports the results for the altermagnetic phase with F-type and ferromagnetic coupling along all directions. The fourth row represents the energy of the magnetic ground state with G-type and ferromagnetic coupling along all directions. }
    \label{TableEnergy}
\end{table*}

\begin{figure}
    
    \includegraphics[width=\linewidth]{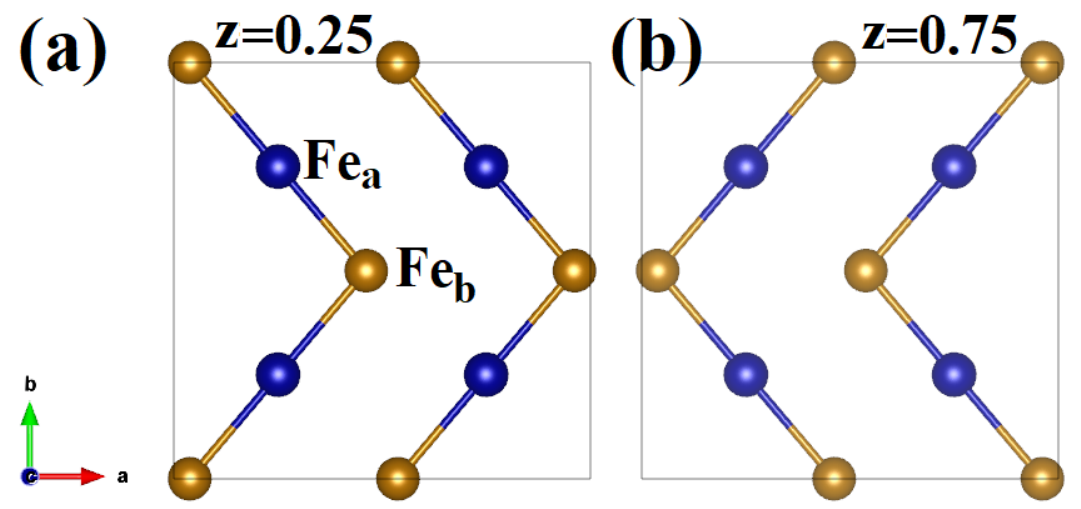}
    \caption{Slice of the zig-zag connectivity of the Fe$_a$ and Fe$_b$ atoms in the $\beta$-Fe$_2$O$_3$ phase. (a) Fe$_a$ and Fe$_b$ atoms at direct coordinate z=0.25. (b) Fe$_a$ and Fe$_b$ atoms at the direct coordinate z=0.75. Given the cubic symmetry, the same connectivity is present for slices in all the equivalent directions. The Fe$_a$ and Fe$_b$ and oxygen atoms are represented by brown and blue, respectively. The oxygen atoms are not shown.}
    \label{connectivity}
\end{figure}

\section{Results and Discussion}

In the next subsections, we will expose the structural properties and the atomic electronic configuration. Later, we will search among the independent magnetic configurations to determine the magnetic ground state of the $\beta$-phase Fe$_2$O$_3$. 

\subsection{Structural properties and atomic electronic configuration}

The $\beta$-phase hosts two kinds of Fe atoms that are surrounded by different octahedra, we will name them Fe$_a$ and Fe$_b$. The unit cell of the $\beta$-phase is composed of 8 Fe$_a$ atoms and 24 Fe$_b$, positioned as depicted in Fig.\ref{crystal} (a). The 8 Fe$_a$ atoms lie on the corner and are in the high-symmetry positions (0.50$\pm$0.25a,0.50$\pm$0.25a,0.50$\pm$0.25a) where a is the lattice constant as shown in Fig. \ref{crystal}(b). The system is cubic therefore all directions are equivalent. We consider the slices of the crystal structure at z=0.25 and z=0.75 which host only Fe atoms. Their positions are represented in Figs. \ref{connectivity}(a) and (b) where we can see two zig-zag chains with a sequence of Fe$_a$-Fe$_b$-Fe$_a$-Fe$_b$ for every slice. Along these chains, the first neighbors are the atoms Fe$_a$ and Fe$_b$ alternated. These chains form an opposite angle at z=0.25 and z=0.75.

From the stoichiometry, we expect that the Fe atoms have a d$^5$ electronic configuration. In our DFT results, all the magnetic moments range from 4.0 to 4.2 $\mu_B$ calculated in a real-space sphere around the Fe atoms. The small variations depend on the kind of Fe atoms (Fe$_a$ or Fe$_b$) and the value of U, however, we can assume from our results that all Fe atoms are always in a high-spin magnetic configuration with S=$\frac{5}{2}$. The magnetic coupling between two atoms with half-filling configuration results in an antiferromagnetic coupling due to the virtual hopping (t) producing a gain in energy proportional to -$\frac{t^2}{U}$ especially in transition metal oxides\cite{doi:10.1021/ic961502+,Ivanov2016}. Due to the structural properties, it is not possible to have a magnetic configuration that satisfies all antiferromagnetic couplings. Therefore, the search for the magnetic ground state is not straightforward.

\begin{figure} 
    \includegraphics[width=0.90\linewidth]{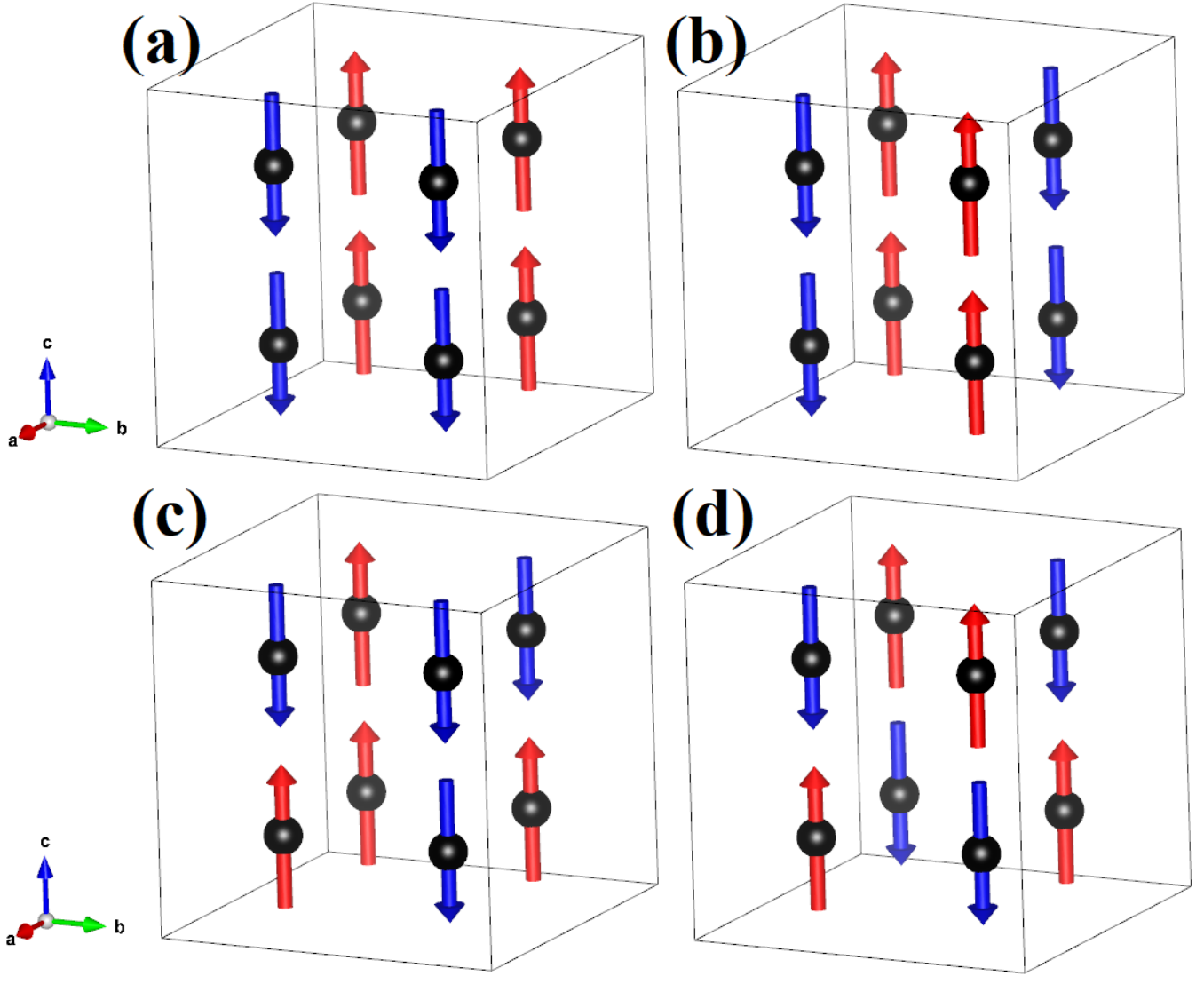}
    \caption{Inequivalent magnetic configurations of the Fe$_a$ atoms with zero net magnetic moment for a cubic system. The antiferromagnetic configurations are usually named (a) A-type, (b) C-type, (c) F-type and (d) G-type. The Fe$_b$ atoms are shown as black balls, while the red and blue arrows represent the spin-up and spin-down, respectively.}
    \label{configuration}
\end{figure}

\subsection{Possible magnetic configurations}

To gain some insight into the magnetic properties, we calculate the total energy as a function of the Coulomb repulsion for the two easiest magnetic configurations: the ferromagnetic phase with all spins aligned (FM) and the ferrimagnetic phase with Fe$_a$ and Fe$_b$ having opposite alignment (FiM). The results are reported in the first two rows of Table I. The ferrimagnetic phase is much lower in energy with respect to the ferromagnetic phase meaning that the first-neighbor magnetic coupling between Fe$_a$ and Fe$_b$ (that we define as J$_{Fe_aFe_b}$) is strongly antiferromagnetic. 
Excluding the energy difference at U=0, where the system is in the unphysical metallic phase, we find that the energy difference decreases as a function of U.
Since the energy difference got reduced as a function of the Coulomb repulsion (but does not change sign), we expect that in lower dimensions as in magnetic nanoparticles, the N\'eel temperature will get reduced too. Therefore, one of the reasons for the low N\'eel temperature of the $\beta$-phase resides in the dimensionality of the experimental samples. The weakening of the magnetic exchange when U increases can be attributed to the magnetic superexchange via virtual hopping, which is the dominant mechanism for the d$^5$-d$^5$ magnetic interaction\cite{doi:10.1021/ic961502+,Ivanov2016}. In this model, J$_{FeFe}$ is proportion to -$\frac{t^2}{U}$, and therefore, it decreases as U increases.

Experimental studies suggest that the ground state is an antiferromagnetic phase. To be antiferromagnetic, the $\beta$-phase should have an antiferromagnetic arrangement within both kinds of octahedra Fe$_a$O$_6$ and Fe$_b$O$_6$.
The equivalent magnetic configurations with zero magnetization for a cubic system with 8 Fe$_a$ atoms are 4 and shown in Fig. \ref{configuration}(a-d)\cite{PhysRev.100.545}. 
For the Fe$_b$ atoms, we can have a very large number of configurations but we assume the ones that are uniform along the zig-zag chain along a given axis. Starting from the spins of the Fe$_a$ atoms, we can have that the first neighbor of the zig-zag chain along the a, b, and c direction can be either FM or AFM and this will keep the system with zero magnetization. Therefore, we can have that along a-, b- and c-axis we have 2 different ferromagnetic couplings (FM or AFM) and we have as possible configuration (FM, FM, FM) while the other inequivalent magnetic configurations are (FM, FM, AFM), (FM, AFM, AFM) and (AFM, AFM, AFM). Therefore, we have 4 inequivalent magnetic configurations for Fe$_a$ and 4 inequivalent magnetic configurations for Fe$_b$. In total, we need to analyze 16 magnetic configurations with zero net magnetization as a function of the Coulombian repulsion.

After the analyses of the energies of all possible magnetic configurations, we determine that the magnetic ground state is composed of the configuration G-type for the Fe$_a$ atoms and the Fe$_b$ atoms being all FM coupled to the Fe$_b$ atoms. Since the G-type magnetism is the ground state, we derive that the magnetic coupling J$_{Fe_a-Fe_a}$ is also antiferromagnetic as expected.
The second lowest energy state is composed of the configuration G-type for the Fe$_a$ atoms and the Fe$_b$ atoms being all (FM,FM,AFM) coupled to the Fe$_b$ atoms. The third lowest energy state is composed of the F-type configuration and the Fe$_b$ atoms being all FM coupled to the Fe$_b$ atoms.
The energies of the magnetic ground state and the energy of the third magnetic ground state are reported in Table I. Even for the magnetic ground state, the energies of the different magnetic phases tend to get reduced but do not change sign as a function of the Coulomb repulsion, meaning that in the case of low dimension the N\'eel temperature decreases due to the increasing of the ratio between U and bandwidth. Additionally, the Mermin-Wagner theorem forbids magnetism in dimensions lower than two. Therefore, we expect that 0-dimensional nanoparticles are even less magnetic. 
The explicit magnetic configuration of the bulk will be shown and discussed in the next Section.

\begin{figure}
\includegraphics[width=0.497\linewidth]{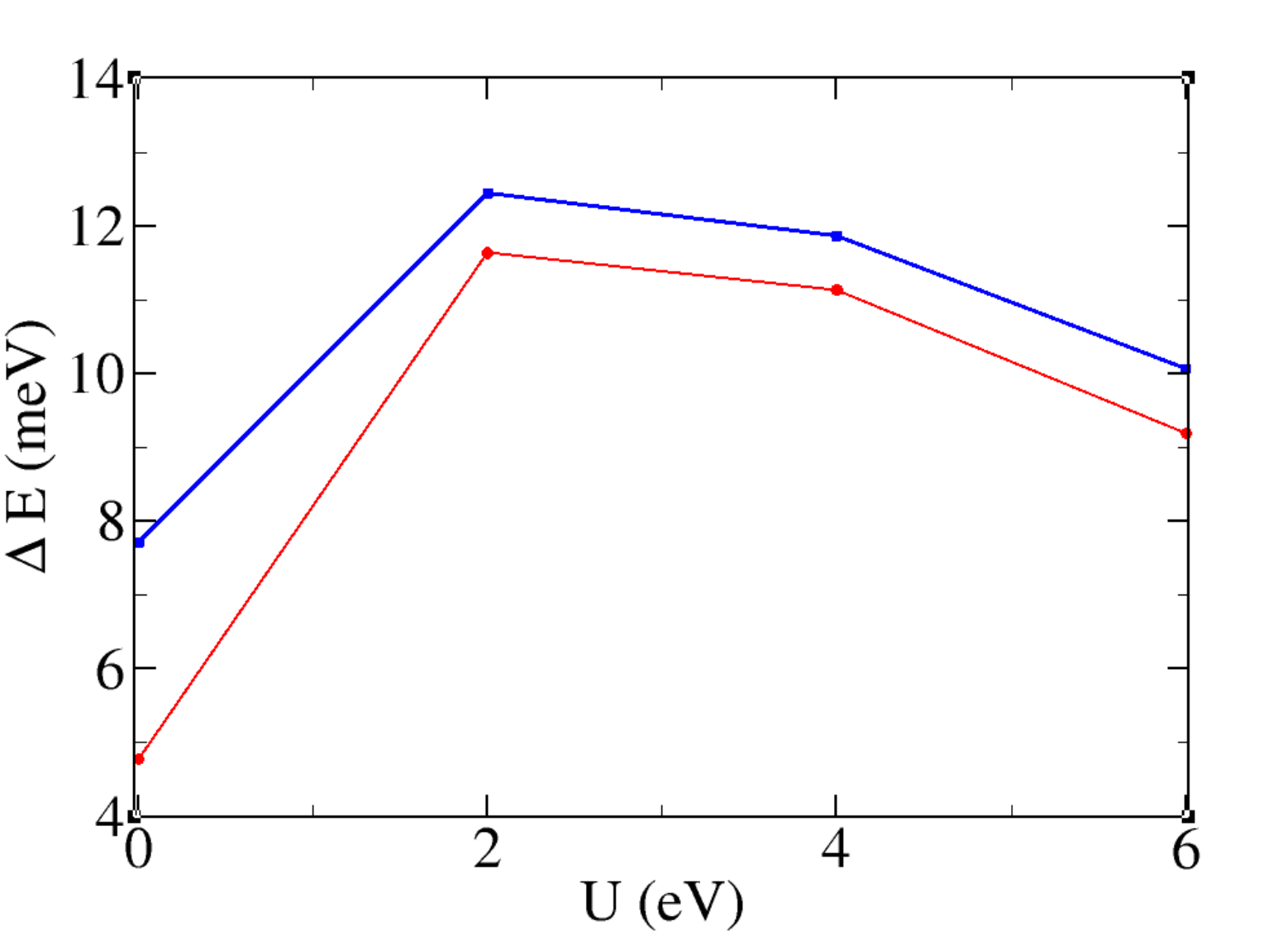}  
        \includegraphics[width=0.497\linewidth]{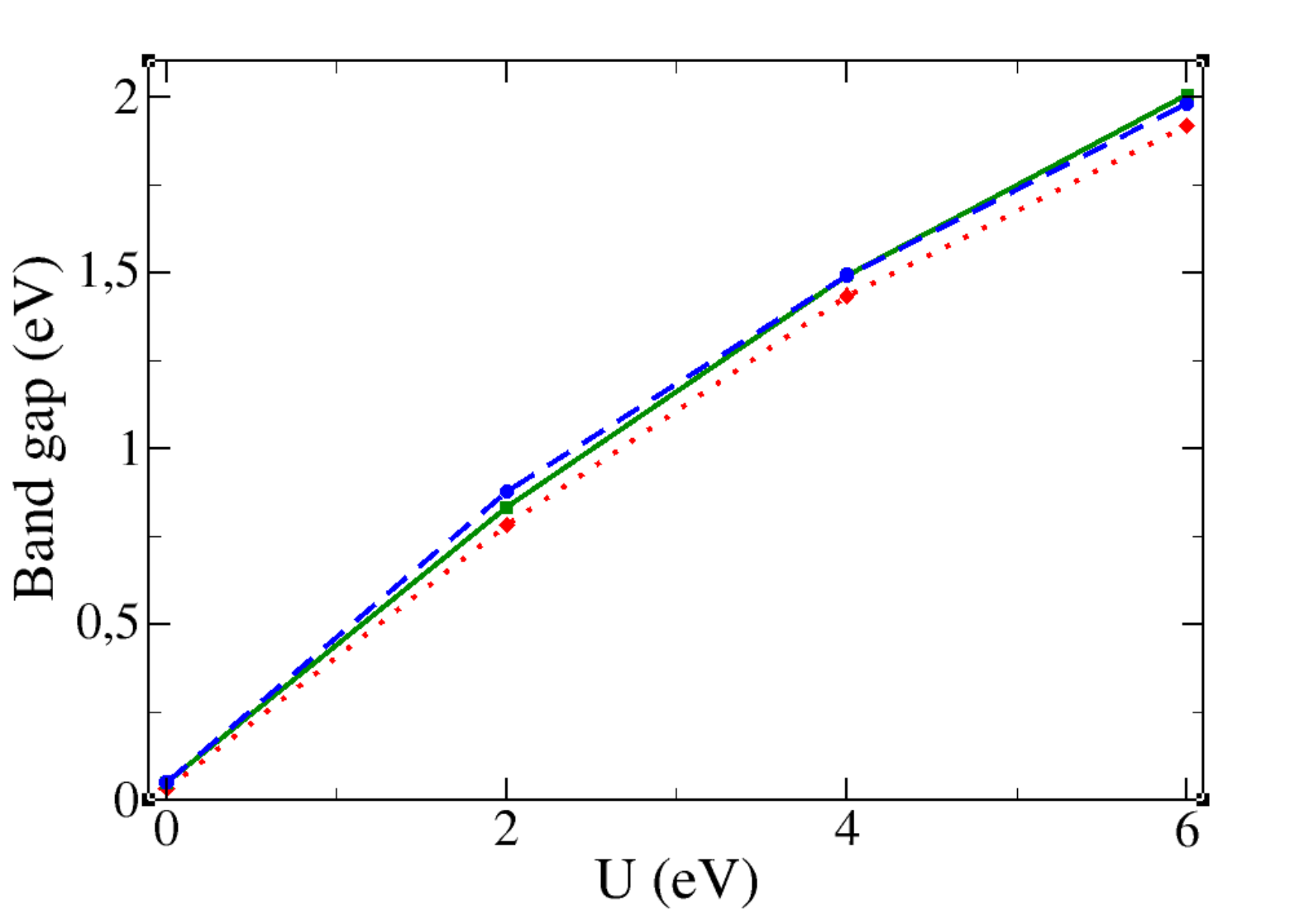}        
        \caption{(a) Energy differences per formula units and (b) evolution of the band gap as a function of the Coulomb repulsion U for the magnetic configurations closer to the ground state of the bulk $\beta$-Fe$_2$O$_3$.  The Coulomb repulsion ranges from 0 to 6 eV. (a) $\Delta$E$_1$ and $\Delta$E$_2$ are represented with solid red and blue, respectively. (b) The gap of the magnetic ground is plotted in dotted green. The second and third magnetic states in energy are plotted in dashed red and solid blue, respectively.}
    \label{Bulk_beta}     
\end{figure} 

\section{Magnetic ground state of the $\beta$-phase}

Once established what is the magnetic ground state, we will calculate the density of the state, the evolution of the gap as a function of U and the band structure. Finally, we will determine if the zero magnetization structure is a Kramers antiferromagnetic, an altermagnet or a ferrimagnet.

\subsection{Electronic properties}

We have analyzed the energy differences as a function of the Coulomb repulsion to check if the ground state is robust. 
Fig. \ref{Bulk_beta}(a) shows the energy difference per formula unit, labeled as $\Delta{E}_1$ and $\Delta{E}_2$, between the ground state and the second- and third-lowest energy states. The data confirm the stability of the ground state across the range of Coulomb repulsion values. Moreover, for U=4 eV which is the value that is more suitable for this compound, the ground state is more stable than at U=0.

\begin{figure}
    \includegraphics[width=1\linewidth,angle=0]{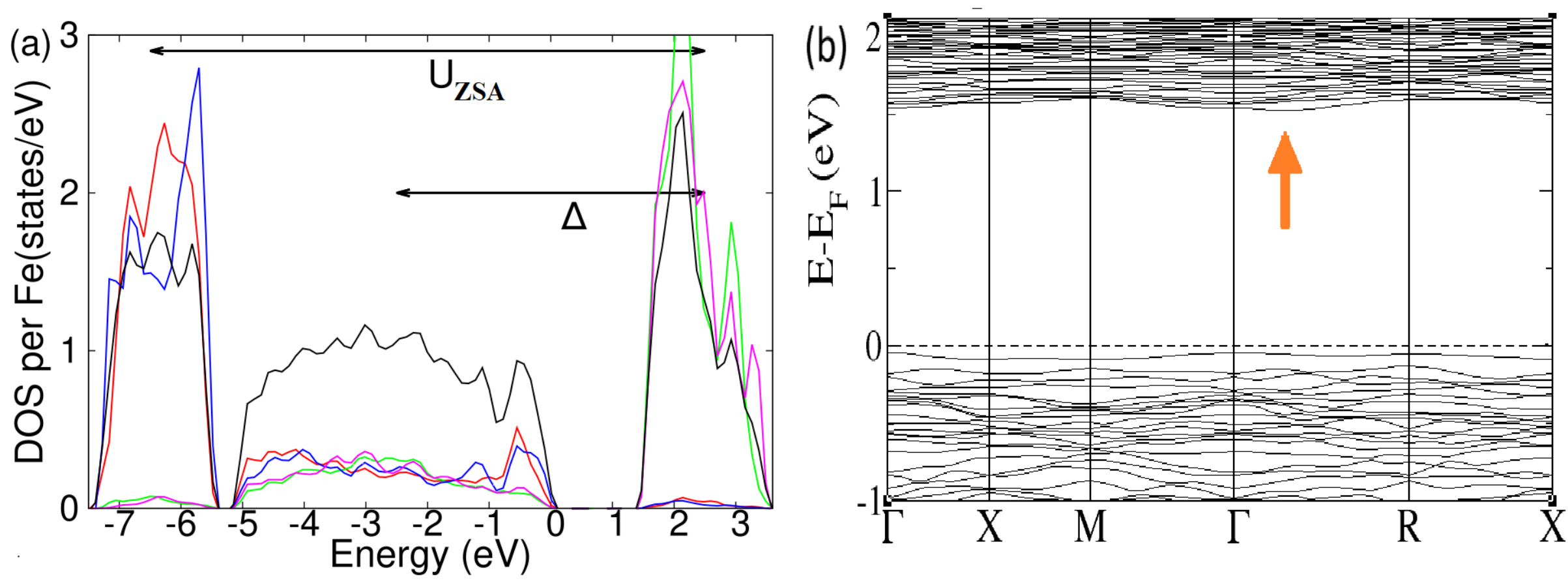}
    \caption{(a) Total and atomic-resolved DOS for the magnetic ground state of the $\beta$-phase. Total DOS per Fe atoms of the Fe$_2$O$_3$ are plotted in black. Majority d-electrons of Fe$_a$ and Fe$_b$ spin are plotted in red and blue, respectively. Minority d-electrons of Fe$_a$ and Fe$_b$ spin are plotted in green and pink, respectively. (b) Band structure of the magnetic ground state with an indirect band gap. The top of the valence band is at the $\Gamma$ point while the bottom of the conduction band is indicated by the orange arrow and placed along the $\Gamma$-R k-path. The band structure is double degenerate.}
    \label{fig:bandstructure}
\end{figure}

Continuing to analyze the electronic properties, we analyze the evolution of the band gap and the density of states. In Fig. \ref{Bulk_beta}(b), the band gap depends linearly on the Coulomb repulsion and it is almost independent of the magnetic configuration. For the realistic value of U=4 eV, the band gap is around 1.5 eV. The band structure of the magnetic ground state for U=4 eV along the high-symmetry k-path is reported in Fig. \ref{fig:bandstructure}(b). Several bands are present due to the large number of atoms in the unit cell, we note that the band gap is indirect since the maximum of the conduction band is at the $\Gamma$ point, while the minimum of the conduction band is along the $\Gamma$-R line indicated by the orange arrow. The atomic-project corresponding DOS is reported in Fig. \ref{fig:bandstructure}(a). The position of the d-electrons levels slightly changes between the two kinds of octahedra Fe$_a$ and Fe$_b$, but we can assume these differences as negligible in the DOS. The bandwidth of the oxygen atoms ranges from the Fermi level to -5 eV. The majority d-electrons lie between -7.5 eV and -5.5 eV, while the minority d-electrons lie between +1.5 eV and +3.5 eV. The system is a charge-transfer insulator since there is a gap between the anions states in the valence band and the cations states in the conduction band\cite{PhysRevLett.55.418}. The charge transfer energy defined as $\Delta$ is around 5 eV for this compound. On the other side, the difference between the majority and minority electrons is defined as U$_{ZSA}$ which is the Coulomb repulsion in the Zaanen-Sawatzky-Allen scheme\cite{PhysRevLett.55.418} and it is equal to 9 eV for this compound. The quantity U$_{ZSA}$ should not be confused with the Coulombian repulsion used within the DFT scheme defined in Section 2.\\  

\begin{figure}
    
    \includegraphics[width=0.345\linewidth,angle=270]{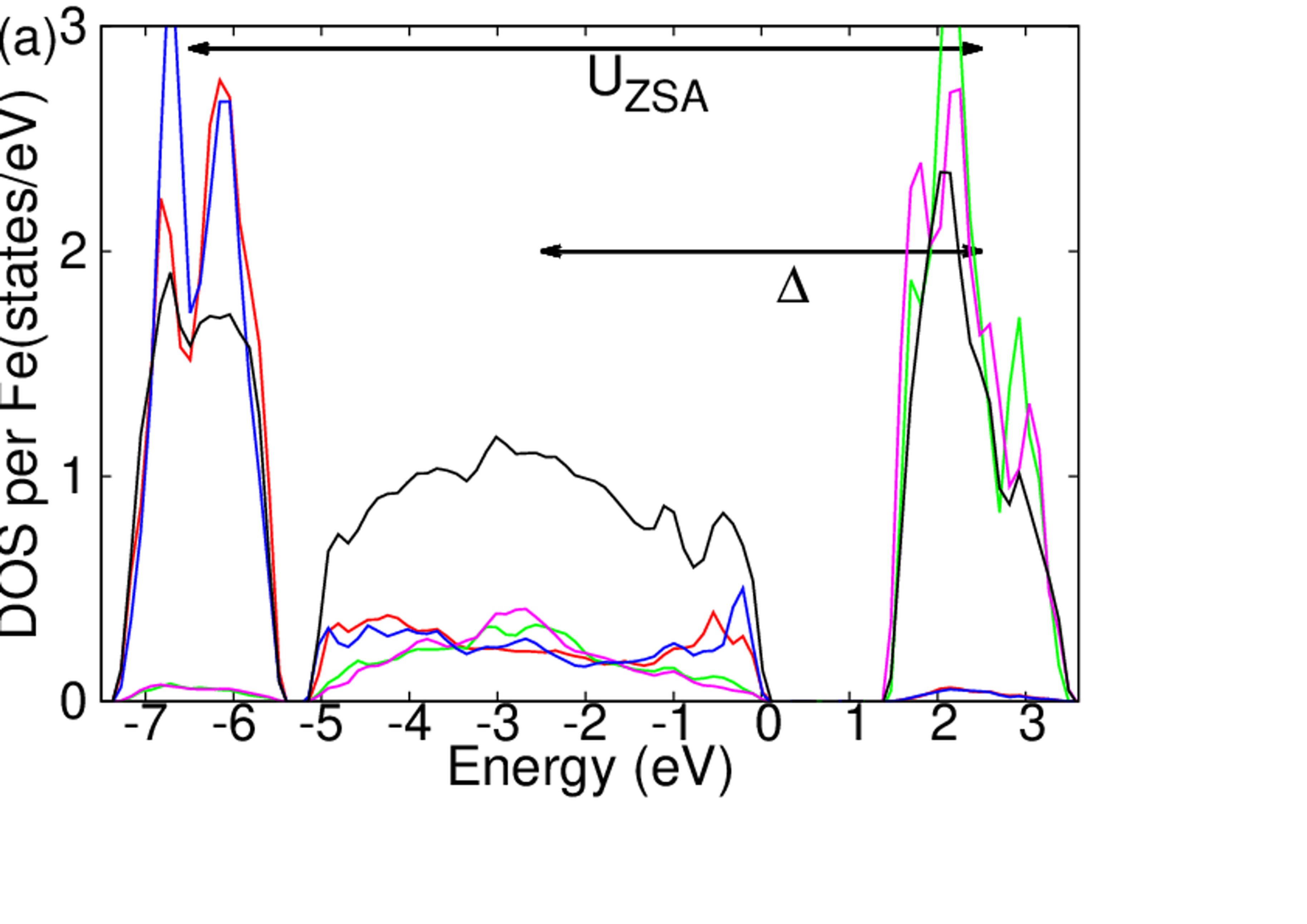}
    \includegraphics[width=0.345\linewidth,angle=270]{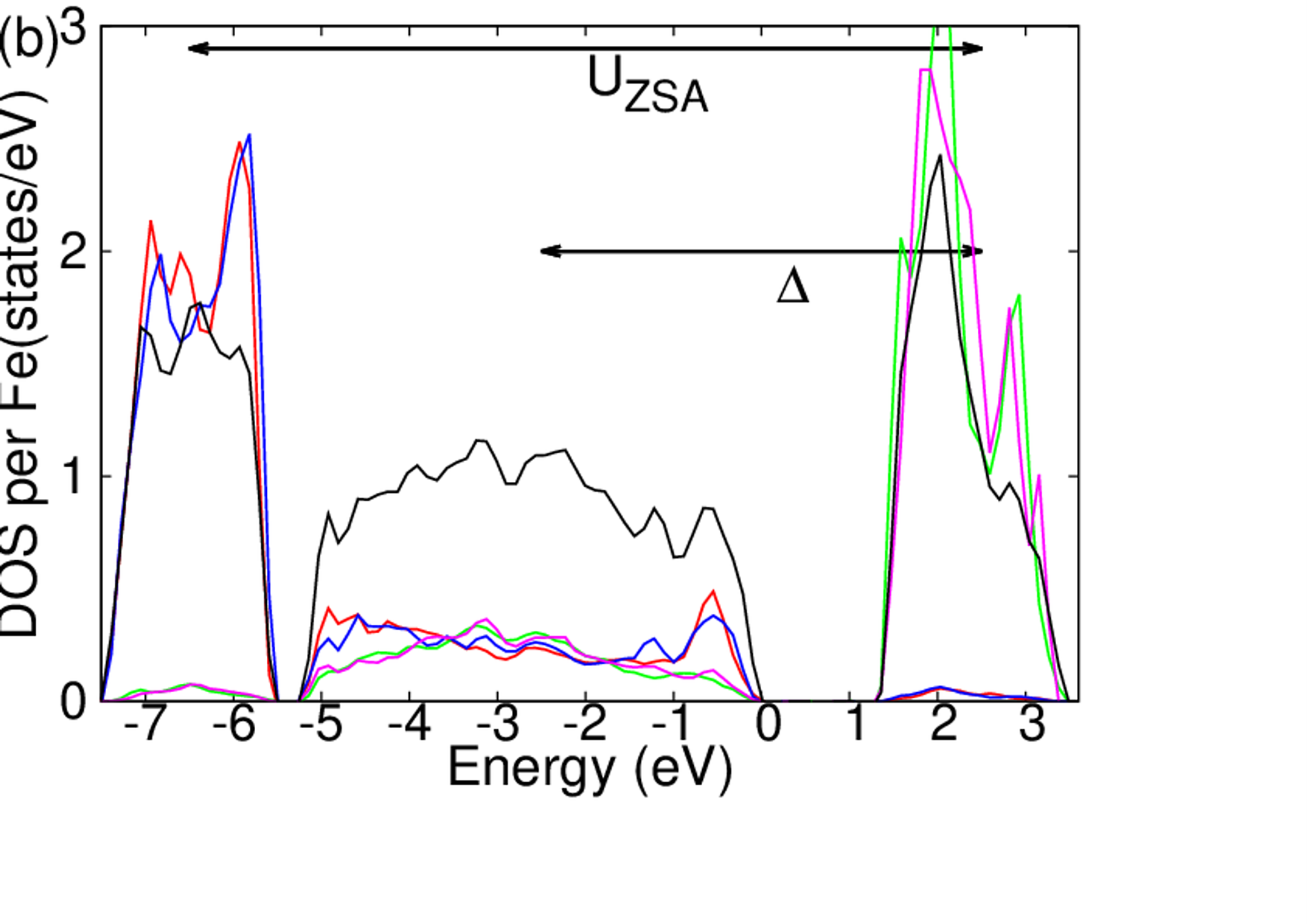}
    \caption{Total and atomic-resolved DOS for the (a) second and (b) third lowest energy state of the $\beta$-phase. Total DOS per Fe atoms of the Fe$_2$O$_3$ are plotted in black. Majority d-electrons of Fe$_a$ and Fe$_b$ spin are plotted in red and blue, respectively. Minority d-electrons of Fe$_a$ and Fe$_b$ spin are plotted in green and pink, respectively. 
    }
    \label{fig:DOS}
\end{figure}

Moving to the other considered magnetic phases, We have calculated the DOSs also for the second and third lowest energy states which are reported in Fig. \ref{fig:DOS}(a) and (b). Beyond minor differences, all the magnetic configurations show an electronic state which is a charge-transfer insulator with similar values of the charge-transfer energy and energy difference between majority and minority d-electrons.

\subsection{Magnetic properties}

\begin{figure}  
    \includegraphics[width=1.0\linewidth,angle=0]{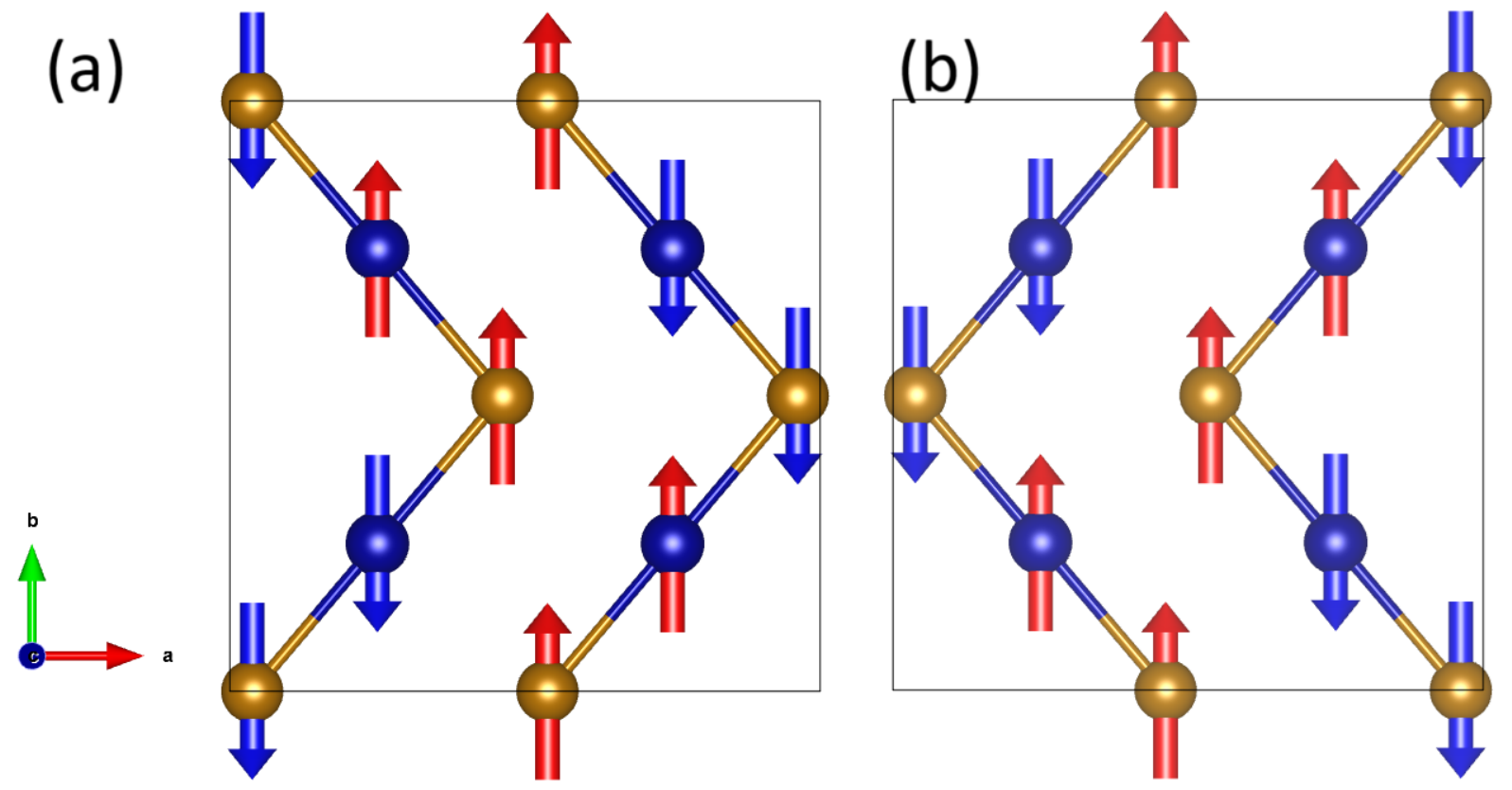}
    \caption{Real space magnetic configuration of the ground state for the slices at (a) z=0.25 and (b) z=0.75. Fe$_a$ and Fe$_b$ atoms are blue and brown, respectively. Spin-up and spin-down are red and blue, respectively.}
    \label{fig:magnetic_configuration}
\end{figure}

The real-space image of the magnetic ground state of the $\beta$-phase is illustrated for the two slices at z=0.25 and z=0.75 in Fig. \ref{fig:magnetic_configuration}(a) and (b), respectively. Along the zig-zag chains, the magnetic configuration is always equivalent to the Fe$_a\uparrow$-Fe$_b\uparrow$-Fe$_a\downarrow$-Fe$_b\downarrow$ magnetic configuration while the spin-up and spin-down alternate moving from a chain to the first-neighbor chain.

The magnetic ground state and the second lowest energy state are Kramers antiferromagnet, while the third ground state is a B-2 d-wave altermagnet state\cite{PhysRevX.12.031042}. Due to the small non-relativistic spin-splitting, we can categorize the latter as fragile altermagnetism\cite{PhysRevMaterials.8.064403} as usually found in transition metal oxides\cite{Cuono23orbital}.

\begin{figure}
    
    \includegraphics[width=0.89\linewidth,angle=0]{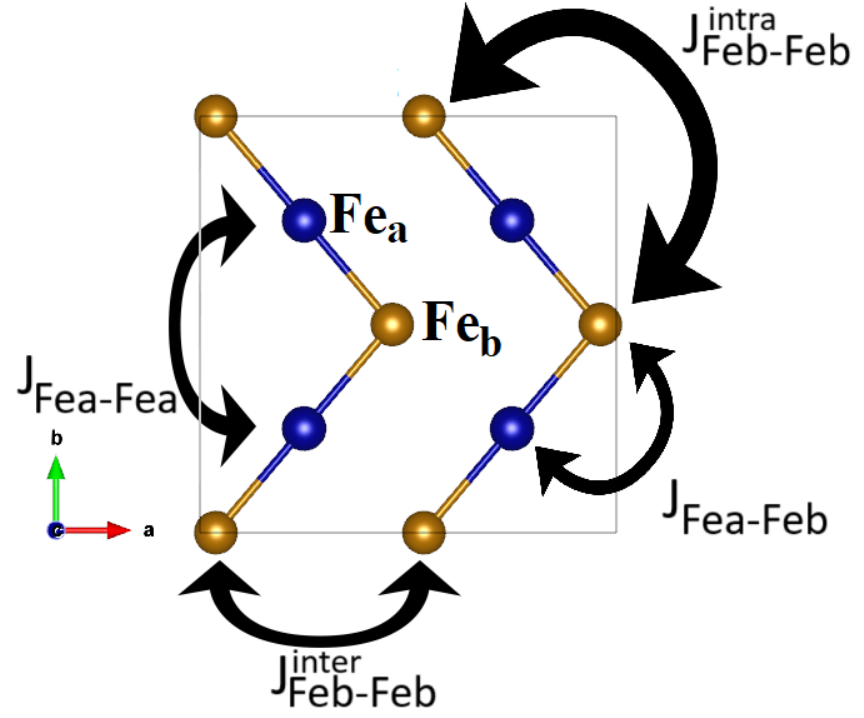}
    \caption{Main exchange couplings for the $\beta$-phase of Fe$_2$O$_3$. The only first neighbor coupling is J$_{Fe_a-Fe_b}$. Regarding the second neighbors couplings, these are between atoms of the same kind which can be interchain (inter) or intrachain (intra). Due to the structural properties, the interchain and intrachain coupling are equivalent for Fe$_a$ atoms, but not for Fe$_b$ atoms. 
    }
    \label{fig:magnetic_coupling}
\end{figure}

The visualization of the relevant magnetic exchanges is reported in Fig. \ref{fig:magnetic_coupling}. The main magnetic couplings are J$_{Fe_a-Fe_a}$, J$_{Fe_b-Fe_b}^{inter}$ and J$_{Fe_b-Fe_b}^{intra}$. 
More in detail, there is one first-neighbor magnetic exchange J$_{Fe_a-Fe_b}$ and three second-neighbor magnetic exchanges which are J$_{Fe_a-Fe_a}$, J$_{Fe_b-Fe_b}^{inter}$ and J$_{Fe_b-Fe_b}^{intra}$. The second-neighbor magnetic exchanges between atoms of the same kind can be interchain (inter) or intrachain (intra). Due to the structural properties, the interchain and intrachain coupling are equivalent for Fe$_a$ atoms, but not for Fe$_b$ atoms. In the ground state arrangement, the second neighbors are all antiferromagnetically coupled and therefore these are satisfied coupling which will stabilize energetically the magnetic phase. On the other hand, the first neighbor coupling J$_{Fe_a-Fe_b}$ is not fully satisfied since it has the configuration of the spins half ferromagnetically and half antiferromagnetically ordered.  
From this information, we infer information on the critical N\'eel temperature. Using the Heisenberg model, we find that the magnetic ground state’s total energy is unaffected by the first-neighbor magnetic exchange J$_{Fe_a-Fe_b}$. As a consequence, the mean-field estimation of the N\'eel temperature will depend exclusively on the second-neighbor magnetic exchanges.\cite{PhysRevB.82.165109} 
More accurate calculations based on Monte Carlo calculation could instead find a weak dependence on the first neighbor exchange, however, This is one of the factors that contribute to explaining the low N\'eel temperature compared to the $\beta$-phase compared to the $\alpha$-phase.\\

\section{Conclusions}

We conducted density functional theory (DFT) calculations to investigate the $\beta$-phase of Fe$_2$O$_3$, focusing on its electronic and magnetic properties. Our analysis explored various possible magnetic configurations of the $\beta$-phase, including ferrimagnets, altermagnets, and Kramers antiferromagnets. Our results indicate that the magnetic ground state has zero net magnetization, aligning with experimental observations. We conclude that the bulk $\beta$-phase of Fe$_2$O$_3$ has an insulating Kramers antiferromagnetic ground state. Additionally, we identified a bulk d-wave altermagnetic phase that lies close in energy to this ground state.
We present the density of states and track the evolution of the band gap as a function of electronic correlations. For suitable values of Coulomb repulsion, the system is classified as a charge-transfer insulator with an indirect band gap. Specifically, for a realistic Coulomb repulsion value of U=4 eV, the band gap is calculated to be 1.5 eV. Analysis of the density of states shows oxygen bands spanning from -5 eV to the Fermi level. The majority Fe bands are observed between -7.5 eV and -5.5 eV, while the majority d-electrons occupy the range from +1.5 eV to +3.5 eV.
Regarding the magnetic properties, the unit cell of the $\beta$-phase consists of 8 Fe$_a$ atoms and 24 Fe$_b$ atoms, for each Fe$_a$ atom, there are 3 Fe$_b$ atoms which are along the three main directions x, y and z. The Fe$_a$ atoms are located at the corners of a cube, and their magnetic ground state forms a G-type structure. The Fe$_b$ atoms exhibit the same spin orientation as their nearest Fe$_a$ neighbors along the x-, y-, and z-directions. The crystal structure features zig-zag chains of Fe$_a$-Fe$_b$-Fe$_a$-Fe$_b$, with alternating spin configurations ($\uparrow$-$\uparrow$-$\downarrow$-$\downarrow$) along all three directions. This magnetic configuration is able to satisfy most of the antiferromagnetic couplings and reduce the degree of magnetic frustration of the system.
When the $\alpha$- and $\beta$-phases coexist, the relative amounts of each phase can be inferred from the weak ferromagnetism. This is because the $\alpha$-phase is an altermagnet, exhibiting weak ferromagnetic behavior, whereas the $\beta$-phase is a Kramers antiferromagnet.

In summary, the polymorphs of Fe$_2$O$_3$ differ not only in structure but also in their magnetic properties. The $\alpha$-phase is an altermagnet, the $\gamma$-phase is a ferrimagnet, and the $\beta$-phase is a Kramers antiferromagnet. Notably, the $\beta$-phase displays charge-transfer insulating behavior with an indirect band gap of 1.5 eV.

\bibliography{biblio}

\begin{thebibliography}{999}

\bibitem[Zhang et~al.(2023)Zhang, Zhao, Huang, Zhao, and Yu]{molecules28186671}
Zhang, Y.; Zhao, M.; Huang, J.; Zhao, N.; Yu, H.
\newblock Controllable Synthesis, Photocatalytic Property, and Mechanism of a Novel POM-Based Direct Z-Scheme Nano-Heterojunction $\alpha$-Fe2O3/P2Mo18.
\newblock {\em Molecules} {\bf 2023}, {\em 28}.

\bibitem[Hua et~al.(2024)Hua, Shah, Ullah, Ullah, and Yuan]{molecules29133082}
Hua, S.; Shah, S.A.; Ullah, N.; Ullah, N.; Yuan, A.
\newblock Synthesis of Fe2O3 Nanorod and NiFe2O4 Nanoparticle Composites on Expired Cotton Fiber Cloth for Enhanced Hydrogen Evolution Reaction.
\newblock {\em Molecules} {\bf 2024}, {\em 29}.

\bibitem[Xue and Wang(2020)]{D0NR02705G}
Xue, Y.; Wang, Y.
\newblock A review of the $\alpha$-Fe2O3 (hematite) nanotube structure: recent advances in synthesis{,} characterization{,} and applications.
\newblock {\em Nanoscale} {\bf 2020}, {\em 12},~10912--10932.
\newblock {\url{https://doi.org/10.1039/D0NR02705G}}.

\bibitem[Agarwal et~al.(2024)Agarwal, Li, Khan, Chia, and Muduli]{PhysRevB.109.224408}
Agarwal, R.; Li, Z.; Khan, K.I.A.; Chia, E.E.M.; Muduli, P.K.
\newblock Efficient terahertz emission from R-plane $\ensuremath{\alpha}{\text{-Fe}}_{2}{\mathrm{O}}_{3}/\mathrm{Pt}$.
\newblock {\em Phys. Rev. B} {\bf 2024}, {\em 109},~224408.
\newblock {\url{https://doi.org/10.1103/PhysRevB.109.224408}}.

\bibitem[Luo and Lyu(2024)]{PhysRevB.109.195203}
Luo, H.; Lyu, S.
\newblock Effects of atomic vibrations on the electronic structure and the photocatalytic efficiency of $\ensuremath{\alpha}{\text{-Fe}}_{2}{\mathrm{O}}_{3}$.
\newblock {\em Phys. Rev. B} {\bf 2024}, {\em 109},~195203.
\newblock {\url{https://doi.org/10.1103/PhysRevB.109.195203}}.

\bibitem[Thakur et~al.(2024)Thakur, Kaur, Kaur, Bhowmik, Han, Singh, Husain, and Sohal]{molecules29194583}
Thakur, R.; Kaur, N.; Kaur, M.; Bhowmik, P.K.; Han, H.; Singh, K.; Husain, F.M.; Sohal, H.S.
\newblock Green Synthesis of Magnetic Fe2O3 Nanoparticle with Chenopodium glaucum L. as Recyclable Heterogeneous Catalyst for One-Pot Reactions and Heavy Metal Adsorption.
\newblock {\em Molecules} {\bf 2024}, {\em 29}.
\newblock {\url{https://doi.org/10.3390/molecules29194583}}.

\bibitem[Dzyaloshinsky(1958)]{DZYALOSHINSKY1958241}
Dzyaloshinsky, I.
\newblock A thermodynamic theory of “weak” ferromagnetism of antiferromagnetics.
\newblock {\em Journal of Physics and Chemistry of Solids} {\bf 1958}, {\em 4},~241--255.
\newblock {\url{https://doi.org/https://doi.org/10.1016/0022-3697(58)90076-3}}.

\bibitem[Galindez-Ruales et~al.(2024)Galindez-Ruales, Šmejkal, Das, Baek, Schmitt, Fuhrmann, Ross, González-Hernández, Rothschild, Sinova, You, Jakob, and Kläui]{galindezruales2024altermagnetismhoppingregime}
Galindez-Ruales, E.F.; Šmejkal, L.; Das, S.; Baek, E.; Schmitt, C.; Fuhrmann, F.; Ross, A.; González-Hernández, R.; Rothschild, A.; Sinova, J.;  et~al.
\newblock Altermagnetism in the hopping regime,  2024,  \href{http://arxiv.org/abs/2310.16907}{{\normalfont [arXiv:cond-mat.mtrl-sci/2310.16907]}}.

\bibitem[Autieri et~al.(2023)Autieri, Sattigeri, Cuono, and Fakhredine]{autieri2023dzyaloshinskii}
Autieri, C.; Sattigeri, R.M.; Cuono, G.; Fakhredine, A.
\newblock Dzyaloshinskii-{Moriya} interaction inducing weak ferromagnetism in centrosymmetric altermagnets and weak ferrimagnetism in noncentrosymmetric altermagnets.
\newblock {\em arXiv preprint arXiv:2312.07678} {\bf 2023}.

\bibitem[Mahfouzi and Kioussis(2022)]{PhysRevB.106.L220404}
Mahfouzi, F.; Kioussis, N.
\newblock Bulk generalized Dzyaloshinskii-Moriya interaction in $\mathcal{PT}$-symmetric antiferromagnets.
\newblock {\em Phys. Rev. B} {\bf 2022}, {\em 106},~L220404.
\newblock {\url{https://doi.org/10.1103/PhysRevB.106.L220404}}.

\bibitem[\ifmmode~\check{S}\else \v{S}\fi{}mejkal et~al.(2022)\ifmmode~\check{S}\else \v{S}\fi{}mejkal, Sinova, and Jungwirth]{Smejkal22}
\ifmmode~\check{S}\else \v{S}\fi{}mejkal, L.; Sinova, J.; Jungwirth, T.
\newblock Emerging Research Landscape of Altermagnetism.
\newblock {\em Phys. Rev. X} {\bf 2022}, {\em 12},~040501.
\newblock {\url{https://doi.org/10.1103/PhysRevX.12.040501}}.

\bibitem[Sattigeri et~al.(2023)Sattigeri, Cuono, and Autieri]{D3NR03681B}
Sattigeri, R.M.; Cuono, G.; Autieri, C.
\newblock Altermagnetic surface states: towards the observation and utilization of altermagnetism in thin films{,} interfaces and topological materials.
\newblock {\em Nanoscale} {\bf 2023}, {\em 15},~16998--17005.
\newblock {\url{https://doi.org/10.1039/D3NR03681B}}.

\bibitem[Cuono et~al.(2023)Cuono, Sattigeri, Skolimowski, and Autieri]{Cuono23orbital}
Cuono, G.; Sattigeri, R.M.; Skolimowski, J.; Autieri, C.
\newblock Orbital-selective altermagnetism and correlation-enhanced spin-splitting in strongly-correlated transition metal oxides.
\newblock {\em Journal of Magnetism and Magnetic Materials} {\bf 2023}, {\em 586},~171163.
\newblock {\url{https://doi.org/https://doi.org/10.1016/j.jmmm.2023.171163}}.

\bibitem[Grzybowski et~al.(2024)Grzybowski, Autieri, Domagala, Krasucki, Kaleta, Kret, Gas, Sawicki, Bo{\.z}ek, Suffczy{\'n}ski, et~al.]{grzybowski2024wurtzite}
Grzybowski, M.J.; Autieri, C.; Domagala, J.; Krasucki, C.; Kaleta, A.; Kret, S.; Gas, K.; Sawicki, M.; Bo{\.z}ek, R.; Suffczy{\'n}ski, J.;  et~al.
\newblock Wurtzite vs. rock-salt MnSe epitaxy: electronic and altermagnetic properties.
\newblock {\em Nanoscale} {\bf 2024}, {\em 16},~6259--6267.

\bibitem[Fakhredine et~al.(2023)Fakhredine, Sattigeri, Cuono, and Autieri]{Fakhredine23}
Fakhredine, A.; Sattigeri, R.M.; Cuono, G.; Autieri, C.
\newblock Interplay between altermagnetism and nonsymmorphic symmetries generating large anomalous Hall conductivity by semi-{Dirac} points induced anticrossings.
\newblock {\em Phys. Rev. B} {\bf 2023}, {\em 108},~115138.
\newblock {\url{https://doi.org/10.1103/PhysRevB.108.115138}}.

\bibitem[Verbeek et~al.(2024)Verbeek, Voderholzer, Schären, Gachnang, Spaldin, and Bhowal]{verbeek2024nonrelativisticferromagnetotriakontadipolarorderspin}
Verbeek, X.H.; Voderholzer, D.; Schären, S.; Gachnang, Y.; Spaldin, N.A.; Bhowal, S.
\newblock Non-relativistic ferromagnetotriakontadipolar order and spin splitting in hematite,  2024,  \href{http://arxiv.org/abs/2405.10675}{{\normalfont [arXiv:cond-mat.str-el/2405.10675]}}.

\bibitem[Bousquet et~al.(2024)Bousquet, Lelièvre-Berna, Qureshi, Soh, Spaldin, Urru, Verbeek, and Weber]{Bousquet_2024}
Bousquet, E.; Lelièvre-Berna, E.; Qureshi, N.; Soh, J.R.; Spaldin, N.A.; Urru, A.; Verbeek, X.H.; Weber, S.F.
\newblock On the sign of the linear magnetoelectric coefficient in Cr2O3.
\newblock {\em Journal of Physics: Condensed Matter} {\bf 2024}, {\em 36},~155701.
\newblock {\url{https://doi.org/10.1088/1361-648X/ad1a59}}.

\bibitem[Pourmadadi et~al.(2022)Pourmadadi, Rahmani, Shamsabadipour, Mahtabian, Ahmadi, Rahdar, and Díez-Pascual]{nano12213873}
Pourmadadi, M.; Rahmani, E.; Shamsabadipour, A.; Mahtabian, S.; Ahmadi, M.; Rahdar, A.; Díez-Pascual, A.M.
\newblock Role of Iron Oxide (Fe2O3) Nanocomposites in Advanced Biomedical Applications: A State-of-the-Art Review.
\newblock {\em Nanomaterials} {\bf 2022}, {\em 12}.

\bibitem[Danno et~al.(2013)Danno, Nakatsuka, Kusano, Asaoka, Nakanishi, Fujii, Ikeda, and Takada]{Danno2013}
Danno, T.; Nakatsuka, D.; Kusano, Y.; Asaoka, H.; Nakanishi, M.; Fujii, T.; Ikeda, Y.; Takada, J.
\newblock Crystal Structure of $\beta$-{Fe}$_2${O}$_3$ and Topotactic Phase Transformation to $\alpha$-Fe$_2$O$_3$.
\newblock {\em Crystal Growth and Design} {\bf 2013}, {\em 13},~770–774.
\newblock {\url{https://doi.org/10.1021/cg301493a}}.

\bibitem[Bafna et~al.(2023)Bafna, Deeba, Gupta, Shrivastava, Kulshrestha, and Jain]{molecules28155722}
Bafna, M.; Deeba, F.; Gupta, A.K.; Shrivastava, K.; Kulshrestha, V.; Jain, A.
\newblock Analysis of Dielectric Parameters of Fe2O3-Doped Polyvinylidene Fluoride/Poly(methyl methacrylate) Blend Composites.
\newblock {\em Molecules} {\bf 2023}, {\em 28}.
\newblock {\url{https://doi.org/10.3390/molecules28155722}}.

\bibitem[Tuček et~al.(2015)Tuček, Machala, Ono, Namai, Yoshikiyo, Imoto, Tokoro, Ohkoshi, and Zbořil]{Tuek2015}
Tuček, J.; Machala, L.; Ono, S.; Namai, A.; Yoshikiyo, M.; Imoto, K.; Tokoro, H.; Ohkoshi, S.i.; Zbořil, R.
\newblock Zeta-Fe2O3 – A new stable polymorph in iron(III) oxide family.
\newblock {\em Scientific Reports} {\bf 2015}, {\em 5}.
\newblock {\url{https://doi.org/10.1038/srep15091}}.

\bibitem[Sakurai et~al.(2009)Sakurai, Namai, Hashimoto, and Ohkoshi]{Sakurai2009}
Sakurai, S.; Namai, A.; Hashimoto, K.; Ohkoshi, S.i.
\newblock First Observation of Phase Transformation of All Four {Fe}$_2${O}$_3$Phases ($\gamma$ $\rightarrow$ $\epsilon$ $\rightarrow$ $\beta$ $\rightarrow$ $\alpha$-Phase).
\newblock {\em Journal of the American Chemical Society} {\bf 2009}, {\em 131},~18299–18303.
\newblock {\url{https://doi.org/10.1021/ja9046069}}.

\bibitem[Hohenberg and Kohn(1964)]{PhysRev.136.B864}
Hohenberg, P.; Kohn, W.
\newblock Inhomogeneous Electron Gas.
\newblock {\em Phys. Rev.} {\bf 1964}, {\em 136},~B864--B871.
\newblock {\url{https://doi.org/10.1103/PhysRev.136.B864}}.

\bibitem[Kresse and Hafner(1993)]{Kresse93}
Kresse, G.; Hafner, J.
\newblock Ab initio molecular dynamics for liquid metals.
\newblock {\em Phys. Rev. B} {\bf 1993}, {\em 47},~558--561.
\newblock {\url{https://doi.org/10.1103/PhysRevB.47.558}}.

\bibitem[Kresse and Furthm\"uller(1996)]{Kresse96b}
Kresse, G.; Furthm\"uller, J.
\newblock Efficient iterative schemes for ab initio total-energy calculations using a plane-wave basis set.
\newblock {\em Phys. Rev. B} {\bf 1996}, {\em 54},~11169--11186.
\newblock {\url{https://doi.org/10.1103/PhysRevB.54.11169}}.

\bibitem[Kresse and Joubert(1999)]{Kresse99}
Kresse, G.; Joubert, D.
\newblock From ultrasoft pseudopotentials to the projector augmented-wave method.
\newblock {\em Phys. Rev. B} {\bf 1999}, {\em 59},~1758--1775.
\newblock {\url{https://doi.org/10.1103/PhysRevB.59.1758}}.

\bibitem[Perdew et~al.(1996)Perdew, Burke, and Ernzerhof]{Perdew96}
Perdew, J.P.; Burke, K.; Ernzerhof, M.
\newblock Generalized Gradient Approximation Made Simple.
\newblock {\em Phys. Rev. Lett.} {\bf 1996}, {\em 77},~3865--3868.
\newblock {\url{https://doi.org/10.1103/PhysRevLett.77.3865}}.

\bibitem[Jain et~al.(2013)Jain, Ong, Hautier, Chen, Richards, Dacek, Cholia, Gunter, Skinner, Ceder, and Persson]{Jain2013}
Jain, A.; Ong, S.P.; Hautier, G.; Chen, W.; Richards, W.D.; Dacek, S.; Cholia, S.; Gunter, D.; Skinner, D.; Ceder, G.;  et~al.
\newblock Commentary: The Materials Project: A materials genome approach to accelerating materials innovation.
\newblock {\em APL Materials} {\bf 2013}, {\em 1}.
\newblock {\url{https://doi.org/10.1063/1.4812323}}.

\bibitem[Liechtenstein et~al.(1995)Liechtenstein, Anisimov, and Zaanen]{PhysRevB.52.R5467}
Liechtenstein, A.I.; Anisimov, V.I.; Zaanen, J.
\newblock Density-functional theory and strong interactions: Orbital ordering in Mott-Hubbard insulators.
\newblock {\em Phys. Rev. B} {\bf 1995}, {\em 52},~R5467--R5470.
\newblock {\url{https://doi.org/10.1103/PhysRevB.52.R5467}}.

\bibitem[Rollmann et~al.(2004)Rollmann, Rohrbach, Entel, and Hafner]{PhysRevB.69.165107}
Rollmann, G.; Rohrbach, A.; Entel, P.; Hafner, J.
\newblock First-principles calculation of the structure and magnetic phases of hematite.
\newblock {\em Phys. Rev. B} {\bf 2004}, {\em 69},~165107.
\newblock {\url{https://doi.org/10.1103/PhysRevB.69.165107}}.

\bibitem[Naveas et~al.(2023)Naveas, Pulido, Marini, Hernández-Montelongo, and Silván]{NAVEAS2023106033}
Naveas, N.; Pulido, R.; Marini, C.; Hernández-Montelongo, J.; Silván, M.M.
\newblock First-principles calculations of hematite ($\alpha$-Fe2O3) by self-consistent DFT+U+V.
\newblock {\em iScience} {\bf 2023}, {\em 26},~106033.
\newblock {\url{https://doi.org/https://doi.org/10.1016/j.isci.2023.106033}}.

\bibitem[Momma and Izumi(2011)]{Momma:db5098}
Momma, K.; Izumi, F.
\newblock {{\it VESTA3} for three-dimensional visualization of crystal, volumetric and morphology data}.
\newblock {\em Journal of Applied Crystallography} {\bf 2011}, {\em 44},~1272--1276.
\newblock {\url{https://doi.org/10.1107/S0021889811038970}}.

\bibitem[Weihe and G{\"u}del(1997)]{doi:10.1021/ic961502+}
Weihe, H.; G{\"u}del, H.U.
\newblock Quantitative Interpretation of the Goodenough-Kanamori Rules: A Critical Analysis.
\newblock {\em Inorganic Chemistry} {\bf 1997}, {\em 36},~3632--3639,  \href{http://arxiv.org/abs/https://doi.org/10.1021/ic961502+}{{\normalfont [https://doi.org/10.1021/ic961502+]}}.
\newblock PMID: 11670054, {\url{https://doi.org/10.1021/ic961502+}}.

\bibitem[Ivanov et~al.(2016)Ivanov, Bush, Stash, Kamentsev, Shkuratov, Kvashnin, Autieri, Di~Marco, Sanyal, Eriksson, Nordblad, and Mathieu]{Ivanov2016}
Ivanov, S.A.; Bush, A.A.; Stash, A.I.; Kamentsev, K.E.; Shkuratov, V.Y.; Kvashnin, Y.O.; Autieri, C.; Di~Marco, I.; Sanyal, B.; Eriksson, O.;  et~al.
\newblock Polar Order and Frustrated Antiferromagnetism in Perovskite Pb2MnWO6 Single Crystals.
\newblock {\em Inorganic Chemistry} {\bf 2016}, {\em 55},~2791--2805.
\newblock {\url{https://doi.org/10.1021/acs.inorgchem.5b02577}}.

\bibitem[Wollan and Koehler(1955)]{PhysRev.100.545}
Wollan, E.O.; Koehler, W.C.
\newblock Neutron Diffraction Study of the Magnetic Properties of the Series of Perovskite-Type Compounds $[(1\ensuremath{-}x)\mathrm{La}, x\mathrm{Ca}]\mathrm{Mn}{\mathrm{O}}_{3}$.
\newblock {\em Phys. Rev.} {\bf 1955}, {\em 100},~545--563.
\newblock {\url{https://doi.org/10.1103/PhysRev.100.545}}.

\bibitem[Zaanen et~al.(1985)Zaanen, Sawatzky, and Allen]{PhysRevLett.55.418}
Zaanen, J.; Sawatzky, G.A.; Allen, J.W.
\newblock Band gaps and electronic structure of transition-metal compounds.
\newblock {\em Phys. Rev. Lett.} {\bf 1985}, {\em 55},~418--421.
\newblock {\url{https://doi.org/10.1103/PhysRevLett.55.418}}.

\bibitem[\ifmmode~\check{S}\else \v{S}\fi{}mejkal et~al.(2022)\ifmmode~\check{S}\else \v{S}\fi{}mejkal, Sinova, and Jungwirth]{PhysRevX.12.031042}
\ifmmode~\check{S}\else \v{S}\fi{}mejkal, L.; Sinova, J.; Jungwirth, T.
\newblock Beyond Conventional Ferromagnetism and Antiferromagnetism: A Phase with Nonrelativistic Spin and Crystal Rotation Symmetry.
\newblock {\em Phys. Rev. X} {\bf 2022}, {\em 12},~031042.
\newblock {\url{https://doi.org/10.1103/PhysRevX.12.031042}}.

\bibitem[Maznichenko et~al.(2024)Maznichenko, Ernst, Maryenko, Dugaev, Sherman, Buczek, Parkin, and Ostanin]{PhysRevMaterials.8.064403}
Maznichenko, I.V.; Ernst, A.; Maryenko, D.; Dugaev, V.K.; Sherman, E.Y.; Buczek, P.; Parkin, S.S.P.; Ostanin, S.
\newblock Fragile altermagnetism and orbital disorder in Mott insulator ${\mathrm{LaTiO}}_{3}$.
\newblock {\em Phys. Rev. Mater.} {\bf 2024}, {\em 8},~064403.
\newblock {\url{https://doi.org/10.1103/PhysRevMaterials.8.064403}}.

\bibitem[Schr\"on et~al.(2010)Schr\"on, R\"odl, and Bechstedt]{PhysRevB.82.165109}
Schr\"on, A.; R\"odl, C.; Bechstedt, F.
\newblock Energetic stability and magnetic properties of MnO in the rocksalt, wurtzite, and zinc-blende structures: Influence of exchange and correlation.
\newblock {\em Phys. Rev. B} {\bf 2010}, {\em 82},~165109.
\newblock {\url{https://doi.org/10.1103/PhysRevB.82.165109}}.

\end{thebibliography}

\end{document}